# A Compact Formula-Aware Spreadsheet Abstraction for Large Language Models

Tie Cheng
SAS Matrix Lead, France
tie.cheng@matrixlead.com

## ABSTRACT

*Large language models (LLMs) are increasingly used to assist spreadsheet users with tasks such as spreadsheet model construction, formula generation, and data analysis, yet their interaction with real-world spreadsheets remains limited by the mismatch between the structure of spreadsheets and the sequential textual representations typically provided to LLMs. We propose a compact formula-aware spreadsheet abstraction that represents large homogeneous ranges through their positions, types, R1C1 formula patterns, and selected boundary values. An efficient algorithm detects and structures these ranges before a textual serialization is supplied to an LLM. The abstraction is provided to the LLM before detailed spreadsheet content is submitted, improving the accuracy and efficiency of spreadsheet processing while reducing token consumption. We evaluate the method on 50 spreadsheet-query cases. The evaluation shows that the abstraction alone was sufficient to answer some queries. When additional spreadsheet content is required, the abstraction reduces the amount of data that must be submitted to the LLM, thereby reducing token consumption. The abstraction layer also improves the reliability and accuracy of spreadsheet processing. We conclude by discussing further techniques for combining spreadsheet static analysis with artificial intelligence to improve the reliability, accuracy, and efficiency of AI-assisted spreadsheet tasks.*

## 1 INTRODUCTION

Spreadsheet software such as Microsoft Excel, Google Sheets, and LibreOffice Calc is widely used for handling data in areas such as engineering, finance, management, statistics, and decision-making. A spreadsheet consists of data and formulas arranged in tables. Advanced users write formulas to automate calculations and build reusable models. Spreadsheets are widely used because their two-dimensional representation of tables is natural for humans, and because they offer a high level of flexibility and power while requiring little or no prior programming knowledge.

Since 2023, with the rapid adoption of generative artificial intelligence, large language models (LLMs), including OpenAI's GPT models, Anthropic's Claude models, and others, have become widely usable by software developers and software publishers. Many innovative tools powered by these models have emerged for spreadsheet users. These include Microsoft 365 Copilot, Gemini in Google Sheets, Shortcut AI, and tools developed by our team, including AI-aided Formula Editor, Spreadsheet AI, and AI Agent for Excel. Such tools aim to assist users with common spreadsheet tasks, including building models, creating formulas, formatting worksheets, and analyzing data. Users can enter requests in natural language, and these tools can perform or suggest actions on their behalf, thereby offering a much higher degree of automation than traditional spreadsheet software has provided until now.

Despite the rapid development of commercial AI tools for spreadsheets, the detailed techniques used to represent or transmit spreadsheet data to LLMs are usually not publicly documented. Academic work has proposed spreadsheet encoding methods for LLMs [Dong et al., 2024], but their adoption in commercial spreadsheet software has not been clearly documented. In this paper, we present and evaluate our technique, which has been integrated into our commercially available software, AI Agent for Excel.

## A Vanilla Approach for Transmitting Spreadsheet Data to LLMs

A straightforward approach for transmitting spreadsheet data to LLMs is to convert spreadsheet ranges into raw textual representations, such as JSON, Markdown, HTML, or lists of cell addresses and values, and submit these representations directly to the model. We refer to this approach as the vanilla approach. Although simple to implement, it is not well suited to large and complex real-world spreadsheets. The main problems are as follows: (1) Results can be incorrect, even for small spreadsheets and simple queries. For example, LLMs may show an imprecise understanding of cells or of the spreadsheet structure. When given a table of 100 rows and 5 columns and asked for the value of the cell located at row 3, column 4, they may return an incorrect answer. (2) The approach can lead to excessive token consumption, making it costly to use on large real-world spreadsheets. (3) It can result in slow response times when applied to large worksheets. (4) Formulas, which are essential for the automation and reuse of spreadsheets, are not handled as effectively as data. Formula manipulation and analysis remain substantially less reliable than data-oriented tasks.

This overall lack of reliability, cost control, efficiency, and power stems from the mismatch between LLMs and the specific characteristics of spreadsheets. On the one hand, spreadsheets have their own complexity: two-dimensional structures, organization into multiple tables, formulas embedded beyond the data, free-form layouts, frequent merged or empty cells, and so on. On the other hand, LLMs were trained mainly on linear, sequential textual data. Moreover, LLMs work relatively well with code written in programming languages because their training relied on a massive volume of freely accessible code. By contrast, formulas, which are encapsulated in spreadsheet files, do not constitute a resource that can be directly exploited for LLM training.

## Adding an Abstraction Layer Before Transmitting Spreadsheet Data to LLMs

To address these limitations, in this paper we propose to construct a compact abstraction representing the key structures and characteristics of the working spreadsheet, including data values, formulas, and large homogeneous ranges. Large homogeneous ranges are represented by their location, data type, formula pattern where applicable, and selected boundary values, rather than by all their cell contents.

We provide the LLM with a textual serialization of this abstraction to give it a higher-level understanding of the spreadsheet before detailed content is submitted. Given a user query, the LLM can first determine whether the abstraction is sufficient to answer the question or whether additional values or formulas are required, in which case only the relevant detailed spreadsheet data is sent. In this way, the abstraction layer can reduce token consumption, improve response times, help the LLM focus on the relevant ranges, and improve the reliability of spreadsheet processing.

In this paper, we make the following contributions. First, we propose a compact, formula-aware abstraction for spreadsheets. Second, we present an efficient algorithm for detecting large homogeneous ranges in spreadsheets, which are central to the construction of this abstraction. Third, we evaluate the method on 50 spreadsheet-query cases and show that it can reduce the amount of spreadsheet data submitted to the LLM while improving the reliability and accuracy of spreadsheet processing. Finally, we discuss further techniques for combining spreadsheet static analysis with artificial intelligence to improve the reliability, accuracy, and efficiency of AI-assisted spreadsheet tasks.

## 2 MOTIVATING EXAMPLE

We have a table located in A1:E910 that presents product sales data and contains the columns "Product", "Date", "Unit price", "Quantity", and "Total price". Columns A to D do not contain formulas; column E contains formulas that multiply column C by column D. For instance, E2 = C2 * D2 and E3 = C3 * D3. We now seek to create a table summarizing all products whose total quantity exceeds 10. This type of data summarization task is very common. In the past, it required the use of a pivot table or the manual construction of formulas. It can now be carried out automatically using artificial intelligence tools, such as the copilot integrated into our product, Spreadsheet AI. The user enters the following command in the conversation box and runs it: "We have a table located in A1:E910. Could you give me the formula to create a summary list of all unique products whose total quantity exceeds 10?" The copilot then returns the following formula: =FILTER(UNIQUE(A2:A910); BYROW(UNIQUE(A2:A910); LAMBDA(x; SUMIFS(D2:D910; A2:A910; x) > 10))) which can then be copied and pasted into a cell to obtain the summary list in question.

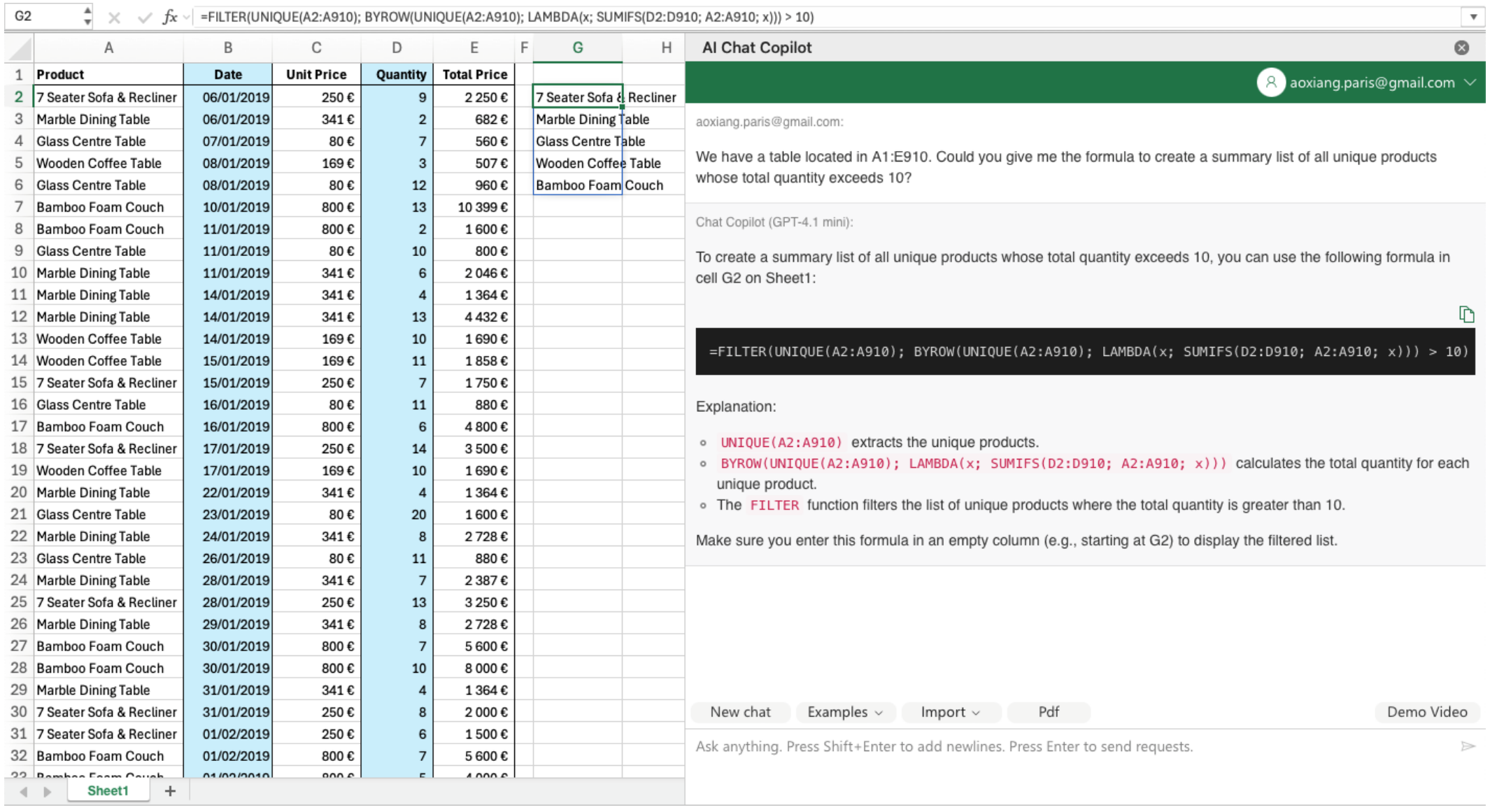

Figure 1. Motivating Example

## The Vanilla Approach

Initially, our tool processes the query through the following steps:

1. It builds a prompt containing the text of the query, together with a question asking which spreadsheet ranges—in terms of values and formulas—are required to answer it, then sends this prompt to an LLM, such as OpenAI's GPT-4.1 (gpt-4.1-2025-04-14).

The user has a worksheet and has the following question. Could you first tell me the values or formulas of which ranges in the worksheet are required to answer it?

We have a table located in A1:E910. Could you give me the formula to create a summary list of all unique products whose total quantity exceeds 10?

2. The model indicates that a relevant answer requires knowing the content of the range A1:E910.
3. The tool then retrieves the values from this range.
4. It builds a new prompt with these values in JSON format, as shown below, and sends it to the model, together with the previous prompt and response in order to preserve the context. Note that the ellipses are used here to simplify our text: in reality, all data from the range is transmitted literally in the message.

Here are the values of A1:E910 of the worksheet that you require. Could you answer the user's question now?

[['Product', 'Date', 'Unit Price', 'Quantity', 'Total Price'],

['7 Seater Sofa & Recliner', '06/01/2019', 250, 9, 2250],

['Marble Dining Table', '06/01/2019', 341, 2, 682],

…, …,

…, …,

['Wooden Coffee Table', '30/12/2020', 169, 8, 1352],

['7 Seater Sofa & Recliner', '30/12/2020', 250, 14, 3500]]

5. The model processes the information received and returns a response to the tool, which displays it to the user.

This processing of a user query through two submissions to the model—steps 1 and 4—is a very common approach. The prompt in step 1 is relatively simple, since it mainly contains the text of the user query accompanied by a question. By contrast, the prompt in step 4 is the real challenge for the tool. This vanilla representation can become very large in the case of large spreadsheets—a frequent situation. This can lead to the following problems. (1) Frequently incorrect results, even for simple queries. For example, these tools often struggle to correctly interpret cells or the structure of the spreadsheet. In the example above, if the model is asked for the value of the cell located at row 10, column 2, it may answer incorrectly. This imprecision is due to the fact that the LLM relies mainly on the general meaning of the text rather than on a rigorous reading of the tabular structure. It is not good at precisely counting commas or parentheses, even though their number is crucial for faithfully representing a table. (2) Rejection of the query by the model if the number of tokens exceeds its maximum capacity. (3) Substantial token consumption even if the query is accepted, increasing costs and processing time. (4) Lack of information about the formulas present in the spreadsheet. Formulas may contain important semantic elements, revealing the meaning of the content and the relationships between cells. Integrating them could not only enable the model to better answer formula-related queries but also improve accuracy for data-related queries and, more broadly, for general queries.

These observations are consistent with the findings of recent studies [Zhang et al., 2024; Li et al., 2024; Sui et al., 2024; Dong and Wang, 2024] on the effectiveness of different spreadsheet representation methods, such as Markdown, HTML, JSON, and DFLoader.

## 3 SPREADSHEET ABSTRACTION FOR LLMS

### Static Analysis of Spreadsheet Applications by Abstract Interpretation

A static analysis technique based on abstract interpretation for verifying spreadsheet properties was introduced by [Cheng and Rival, 2012; Cheng and Rival, 2015]. This approach starts from the observation that the data contained in a spreadsheet may evolve over time. For example, users may update entered data on a daily basis. Moreover, dynamic array functions such as UNIQUE and FILTER can change the dimensions of the resulting table depending on the input values. Additionally, data and formulas can be set and manipulated by VBA (Visual Basic for Applications) programs. A rigorous and reliable analysis must therefore consider all possible concrete states of the spreadsheet when inputs are unknown or vary over time, rather than only its current state.

To achieve this, the set of possible concrete states is over-approximated by abstractions, and the analysis reasons over these abstractions. More precisely, the spreadsheet semantics is lifted from concrete states to abstract states: formulas, user inputs, dynamic array computations, and VBA updates are interpreted as abstract transfer functions. These functions are then iteratively applied until a sound post-fixpoint is reached. The resulting abstract state represents a stable over-approximation of all spreadsheet states that may arise during execution or use.

For example, the concrete values of worksheet cells—such as “abc”, 20/07/2025, or 123—can be abstracted by their types: String, Date, or Number. Similarly, concrete formulas can be abstracted: for instance, “=123+345” can be represented as “=Number+Number”. A non-empty range starting from A1, whose number of rows may vary, can be abstracted as [1,1] × [i,1], where i is strictly greater than 0. Finally, a worksheet containing a simple one-column table of numbers starting at A1 can be represented as “[1,1] × [i,1], where i > 0, with Numbers”.

On this basis, a formula—whether entered by the user, proposed by an AI system, or inserted by a VBA program—such as ‘= “abc” + A15’ can be flagged, because it represents the addition of a string to a cell that may contain a number. Such an operation is semantically invalid and may indicate an error in the formula or an anomaly in the spreadsheet. This is the principle underlying type-based spreadsheet correctness verification.

### Spreadsheet Abstraction for LLMs

As discussed in Section 2, the vanilla representations of spreadsheets for LLMs have led to a lack of reliability, cost control, and efficiency. It therefore seemed natural to adopt our abstraction-based approach to optimize this procedure. We construct a spreadsheet abstraction that captures the worksheet’s key high-level characteristics. We then generate a concrete textual serialization of this abstraction and include it in the first request sent to the LLM. This textual serialization provides the model with a high-level view of the worksheet and helps it reason about the worksheet more effectively. To build this abstraction, the use of types to represent cell values, as well as the abstraction of formulas, remains relevant. We therefore designed and implemented the following optimized approach:

1. Given a worksheet, we scan the values and formulas of its cells.
2. For values, we study their type: Text, Date, Number, Currency—for example, “520 €”—and so on.
3. For formulas, we study their representation in R1C1 notation. For example, if cell E2 contains the formula =C2*D2 in A1 notation, its R1C1 notation is =RC[-1]*RC[-2].
4. We then construct the spreadsheet abstraction by partitioning the set of non-empty cells in the spreadsheet into non-overlapping ranges and individual cells, based on the above-mentioned abstractions of values and formulas. Each range consists of cells sharing a single data type and, where applicable, the same formula in R1C1 notation.
5. We generate a textual serialization of the spreadsheet abstraction from these identified cells and ranges.
6. For a cell, we directly provide its value and formula, if any.
7. For a range, we indicate the unique type of its values and, where applicable, its unique formula. We also provide the values of the cells located on the four edges of the range.
8. We send this textual serialization to the LLM, together with the user’s query, to ask it which spreadsheet ranges—in terms of values and formulas—are required to answer the query.
9. The LLM may indicate that the textual serialization already contains sufficient information and that no additional values or formulas are required. In that case, we rerun the user’s query as is, with the previous prompt and response to maintain context, and directly retrieve the response. Otherwise, we send it the precise values and formulas of the identified ranges so that it can answer the user’s query.

Taking the motivating example in Section 2, here is the message we build through steps 1–3 and send to the LLM at step 4. Note that the ellipses are an integral part of the transmitted message; they are not added here merely to simplify the explanatory text.

The user has a worksheet. The following text is a textual serialization of the worksheet abstraction. The worksheet is divided into distinct, non-overlapping cells and ranges. Each part is separated by a blank line and describes either a single cell or a range. For a range, the data type is specified. Each range contains only one type of data, and the values of its edge cells are provided. For a single cell, its value is given directly. If a cell or range includes a formula—only one formula is allowed—it is written in R1C1 format rather than A1 format.

A1: 'Product'

A2:A910: Text

A2: '7 Seater Sofa & Recliner', A3: 'Marble Dining Table', …, A909: 'Wooden Coffee Table', A910: '7 Seater Sofa & Recliner'

B1: 'Date'

B2:B910: Date

B2: '2019-01-06', B3: '2019-01-06', …, B909: '2020-12-30', B910: '2020-12-30'

C1: 'Unit price'

C2:C910: Currency

C2: 250, C3: 341, …, C909: 169, C910: 250

D1: 'Quantity'

D2:D910: Number

D2: 9, D3: 2, …, D909: 8, D910: 14

E1: 'Total price”'

E2:E910: Currency

E2: 2250, E3: 682, …, E909: 1352, E910: 3500

E2:E910’s formula: =RC[-1]*RC[-2]

The user has the following question. Could you first tell me the values or formulas of which ranges in the worksheet are required to answer it?

We have a table located in A1:E910. Could you give me the formula to create a summary list of all unique products whose total quantity exceeds 10?

In this example, the LLM responds that the textual serialization already contains sufficient information to provide the formula requested by the user. We then only need to rerun the user’s query to obtain the exact formula. If the initial query asks for a concrete list rather than a formula—for example: “We have a table located in A1:E910. Could you give me the summary list of all unique products whose total quantity exceeds 10?”—then, at the end of step 4, the LLM will indicate that the values of ranges A1:A910 and D1:D910 are required, rather than the values of the entire range A1:E910. We then transmit these values to the LLM in full and ask it to provide a final answer.

### The Algorithm for Generating the Abstraction

Given a spreadsheet, there are many ways to divide it into cells and ranges. We seek to optimize steps 1 to 3 above in order to build an abstraction that best reflects the structure of the spreadsheet and enables the LLM to grasp its characteristics more effectively. Specifically, a good abstraction—as well as the algorithm responsible for generating it—should meet the following criteria:

- The algorithm must be able to identify the main tables of a spreadsheet so that the abstraction can represent them to the LLM. This includes large tables—for example, those exceeding 10 rows—and main tables, meaning those that capture as broad a structure as possible while remaining coherent, without being limited to narrower subsets. For example, the range A2:E910 constitutes the main large table of the example spreadsheet. Such a table contains several large homogeneous vertical lists, without necessarily being homogeneous itself. A vertical list is said to be large and homogeneous if it uses a single R1C1 formula for all its cells, such as E2:E910, or if it contains a single data type, such as A1:A910 or B2:B910. These lists are then represented by their location, their value type, any single formula, and the values of the cells located at the top and bottom—rather than by all the values of the list.

- The algorithm must be able to detect whether a large table has headers, such as A1:E1, in the spreadsheet. If so, the textual serialization must include the concrete values of these headers, rather than only their types, because those values provide valuable information for understanding the table.

- The textual serialization must present the headers and data of a large table in a coherent and structured manner so that the LLM can more easily grasp its logic. For example, in the example above, it first presents the header of A2:A910, then A2:A910, then the header of B2:B910, then B2:B910, and so on.

- The algorithm must tolerate a certain level of noise in the data. In particular, the presence of empty cells in a well-structured large table should not lead to unjustified fragmentation of that table. The algorithm must be able to "ignore" these empty cells where relevant.

- The generation algorithm must remain inexpensive in time and space, so that it can be applied at scale without degrading the overall performance of the system.

We therefore design the following algorithm. The intuition is to start from the most reliable large homogeneous ranges in the spreadsheet and then progressively extend them into a coherent abstraction. In this sense, the table-construction phase can be viewed as a spreadsheet-specific form of region growing. The algorithm first identifies seed structures from repeated R1C1 formula patterns, since such formulas reveal regular computational structures, such as calculated columns, and also indicate the ranges on which they depend. For example, in the case presented above, the range E2:E910 contains a formula of the form =RC[-1]*RC[-2], which implies not only that it is a homogeneous list, but also that the two lists on which it depends—C2:C910 and D2:D910—are themselves homogeneous lists. Such lists are detected repeatedly by following formula dependencies until no new large list is found. The algorithm then detects another class of reliable large homogeneous ranges: ranges containing a single data type. Subsequently, it expands the set of detected structures by grouping neighboring homogeneous lists with compatible height and position, while tolerating a limited number of empty cells. This process continues until no new list or table extension can be discovered. The algorithm then detects headers and separates them from the data. Finally, it generates the textual serialization by summarizing large homogeneous ranges according to the spreadsheet abstraction and by explicitly recording irregular cells that do not belong to them. More precisely,

1. We retrieve the values and R1C1 formulas of all cells in the spreadsheet.
2. We analyze the R1C1 formulas of the cells, column by column and from top to bottom, in order to identify large vertical lists, for example E2:E910.
3. For each identified large list, we analyze the ranges referenced in its formula—for example, C2:C910 and D2:D910—and then identify, in turn, any large lists among them, for example C2:C910 and D2:D910. This step is repeated until no new large list is detected.
4. At this stage, several reliable large vertical lists have been identified.
5. We analyze the type of spreadsheet cells and identify other large homogeneous vertical lists containing a single data type, for example A1:A910 or B2:B910.
6. We then treat the identified large lists as tables, which we scan and attempt to group in order to form the main tables. For each table, we examine its direct neighbors to the left and right:
    a. If another table has a similar height and a close horizontal position, we group it with the current table.
    b. If a data list has a similar height and a close horizontal position, while tolerating a few empty cells, we also integrate it into the same table.
7. At this stage, several main tables of the spreadsheet have already been identified. For example, A1:A910, B2:B910, C2:C910, D2:D910, and E2:E910 have been grouped into a single main table.
8. We scan all main tables to detect, where applicable, the presence of headers, whether they are located inside or immediately above the table.
    a. If a non-empty row spanning the width of the table is located above a large vertical list with a formula, we consider it a header row. For example, A1:E1 is non-empty and located above E2:E910, so A1:E1 is identified as the table headers.
        i. We draw a separation line between the headers and the data. For each list in the table:
            1. If the list extends beyond this line, we consider the cell extending beyond it to be its header, then reduce the height of the list accordingly. For example, A1 is identified as the header of A1:A910, and the list becomes A2:A910.

2. If the list does not extend beyond this line, we associate the header with the cell located immediately above the list. For example, B1 is identified as the header of B2:B910.

  b. If a non-empty row spanning the width of the table is located above a large vertical list with a different type, we apply the same treatment.

9. Finally, we build the textual serialization:

  a. For each main table, we scan its lists:

    i. If the table contains headers, we display the position and value of the header, followed by the position of the list, its data type, its formula if any, and a few values located at the edges.

    ii. If the table does not contain headers, we directly display the position, data type, formula if any, and a few values at the edges.

  b. For cells that are not part of any main table, we display their position, value, and formula if any, one by one.

## 4 EVALUATION

### Evaluation Case Set

We evaluate the reliability, gain, and efficiency of the spreadsheet abstraction approach for LLMs—hereinafter the "abstraction approach"—by comparing it with the vanilla approach. The vanilla approach corresponds to the implementation used in our tool AI-aided Formula Editor, while the abstraction approach corresponds to the implementation used in our tool AI Agent for Excel. In both cases, the underlying LLM is OpenAI's GPT-4.1 (gpt-4.1-2025-04-14), so that the comparison focuses on the effect of the spreadsheet representation strategy rather than on differences between models.

We built a set of 50 evaluation cases, based on our own case database as well as on real tasks from our pilot users, who agreed to share their queries and data with us. Each test typically consists of a worksheet and a query. A worksheet may contain one or more tables, and its size varies from 100 to 10,000 cells. Query length ranges from a single sentence to several paragraphs. The objectives of the queries include, for example, formula generation, as in the main example of this paper, formula optimization, formula correction, data consolidation, data cleaning, data consistency verification, data prediction, data completion, and data analysis. The evaluation queries can be found under the "Examples" menu in AI Agent for Excel, as shown in Figure 2.

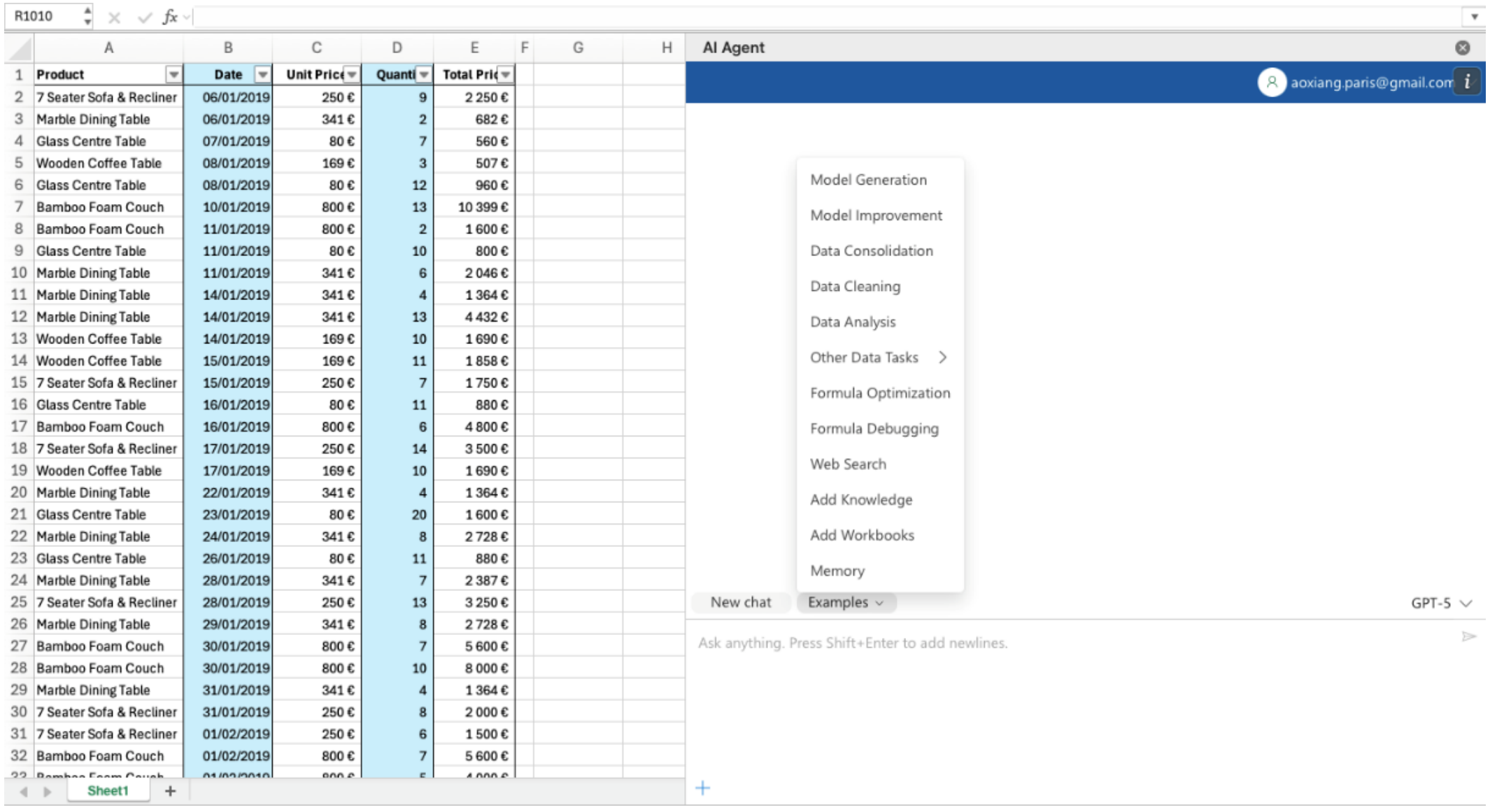

Figure 2. Query categories available in AI Agent for Excel

**Evaluation Results**

**Can our algorithm generate an abstraction that properly represents the characteristics of a spreadsheet, especially its tables?** To verify this, we manually examine the textual serializations generated by the algorithm for all evaluation cases. We confirm that they correctly identify the main large tables of the spreadsheets, as well as their headers when present, and present them in a coherent and structured manner. In addition, cells and ranges located outside the main large tables are also presented in an intuitive and logical way.

**Does spreadsheet abstraction already make it possible to answer queries without requiring access to the spreadsheet content?** This question was evaluated using a single-run protocol. Each evaluation case was executed once with the vanilla approach and once with the abstraction approach. Of the 50 evaluation cases, 12 could be answered correctly without access to any spreadsheet content. Among the remaining 38 cases, the abstraction approach was able to answer 11 correctly using only the textual serialization of the spreadsheet abstraction, without requiring any additional detailed spreadsheet content. By contrast, the vanilla approach still required access—often large-scale access—to the contents of worksheet ranges, which had to be included in a second request to the LLM. This therefore leads to a substantial reduction in query processing time as well as token savings. These cases relate in particular to formula generation, optimization, correction, and analysis.

**Does spreadsheet abstraction also make it possible to better target the relevant ranges?** This question was also evaluated using the same single-run protocol. For each case in which additional spreadsheet content had to be consulted, we recorded the ranges requested by the LLM under both the vanilla approach and the abstraction approach. We then compared the number of cells covered by these ranges. In addition, we measured the total token consumption across the complete interaction, including both the range-identification request and any subsequent request used to obtain the final answer. Among the 27 cases where additional spreadsheet content was required, the abstraction approach identified substantially more precise ranges than those generally requested by the vanilla approach. It reduced the volume of requested cells by 77% and total token consumption across all requests by 86%, resulting in a substantial reduction in query processing time. These cases relate in particular to data analysis.

**Does spreadsheet abstraction make the LLM more reliable in processing spreadsheets?** This question was evaluated using a repeated-run protocol, because reliability concerns the stability of the model's answers across runs. Each case was run three times under the same experimental conditions with both the vanilla approach and the abstraction approach. Each run was manually evaluated and counted as successful if its final answer was correct. Among the 50 evaluation cases, we identified 22 cases in which the vanilla approach failed in at least one of the three runs. The results show that providing the LLM with the textual serialization of the spreadsheet abstraction—which represents the structure of the spreadsheet at a high level—improves the LLM's accuracy in locating cell content and therefore improves the reliability of its answers. In these 22 cases, the abstraction approach substantially improves reliability overall; in 16 cases, it reaches a 100% success rate, meaning that all three runs produced correct answers.

## 5 CONTRIBUTION TO THE STATE OF THE ART

Table detection in a spreadsheet is a fundamental task for spreadsheet understanding. Several studies have been conducted on this subject [Dong et al., 2019; Christodoulakis et al., 2020; Vitagliano et al., 2022]. As part of our static analysis approach based on abstract interpretation aimed at verifying various spreadsheet properties [Cheng and Rival, 2012; Cheng and Rival, 2015], we also designed and implemented an algorithm for identifying homogeneous rectangular zones. Unlike previous work, whose objective was generally table structuring or extraction for human analysis or indexing purposes, the algorithm we designed and implemented in this paper specifically aims to generate an optimized and descriptive representation intended to provide LLMs with an understanding of the spreadsheet. Optimized for this use case, it has demonstrated its effectiveness in this context, thereby making an original contribution to the state of the art:

- We study and optimize in particular the detection and representation of large main tables, because without optimization, transmitting them to LLMs would be costly.
- We leverage the semantics of formulas in the process of detecting these tables—an aspect often underexploited in previous work, which focuses mainly on data.
- To detect large main tables, our method consists in scanning large homogeneous vertical lists and extending the table to the left or right whenever a vertical neighbor has a compatible structure. This method, both intuitive and efficient, also enables robust handling of noise present in the data.

Furthermore, a spreadsheet encoding method for LLMs was proposed at the end of 2024 [Dong et al., 2024] to improve the quality of model responses. Their approach relies on the values contained in worksheets and does not exploit formulas, whereas our abstraction uses formulas not only as seeds for detecting homogeneous ranges, but also as content that enables LLMs to better perform formula analysis—a frequent and important user need.

While our method provides a spreadsheet-structure-based scaffold for LLM-assisted spreadsheet processing, other forms of scaffolding have also been proposed. For instance, Test-Driven Development has been proposed as a cognitive scaffold for LLM-assisted spreadsheet code generation [Thorne and Sarkar, 2025]. Rather than relying on unconstrained natural-language prompting, users first express expected behaviors and test cases, thereby constraining the generation process and supporting the reliability, verifiability, and comprehensibility of generated spreadsheet code.

Early EuSpRIG studies showed that ChatGPT could produce useful spreadsheet-related answers, but also plausible incorrect formulas, false statements, and hallucinations, especially when tasks involved uncertainty, inference, or complex reasoning [O'Beirne, 2023; Thorne, 2023]. More recently, FLARE was proposed as a benchmark showing that recent LLMs perform well on straightforward spreadsheet tasks but remain unreliable on complex, multi-step operations [Thorne, 2025]. While these studies are not specifically concerned with the scale and structure of spreadsheet data, our work focuses on improving the reliability and efficiency of LLM-assisted processing for large, two-dimensional spreadsheets.

## 6 DISCUSSION AND CONCLUSION

### Limitations and Future Work

The algorithm described in this paper detects and abstracts large homogeneous ranges by grouping large homogeneous vertical lists. However, real-world spreadsheets may also contain large homogeneous horizontal lists, although these are less common than vertical lists. A natural extension of the algorithm would therefore be to first detect and abstract large homogeneous ranges by grouping large homogeneous vertical lists, and then to apply an analogous procedure to detect and abstract large homogeneous horizontal lists in the remaining areas of the spreadsheet.

Irregular spreadsheets, in which data cells are dispersed across the worksheet and do not form large homogeneous ranges, may also exist. Our algorithm does not abstract such dispersed cells and instead sends them directly to the LLM. When the number of such cells is small, this is acceptable. However, when the number of dispersed cells is large, this may still lead to substantial token consumption. Future work could address this limitation by introducing additional types of abstraction for sparse regions, for example based on their semantic content. Such an extension would allow the spreadsheet abstraction to combine abstractions of large homogeneous ranges, abstractions of sparse regions, and the remaining concrete cells.

The algorithm described in this paper does not explicitly handle merged cells, although merged cells are common in real-world spreadsheets. This may affect the result of the algorithm, for example when merged cells are used to represent table headers, which may lead to inaccurate header detection. Future work could address this issue by adding a preprocessing step that detects merged regions, represents each merged area explicitly in the spreadsheet abstraction, and propagates its logical value to the cells it covers when this is relevant for table and header detection.

In addition to cell values and formulas on which this paper focuses, spreadsheet formatting can also encode important structural or semantic information. For example, bold fonts, background colors, borders, indentation, and number formats may indicate headers, subtotals, input cells, calculated cells, section boundaries, or cells requiring special attention. Since the current algorithm does not systematically exploit such formatting information, it may miss part of the meaning conveyed by the spreadsheet layout. Future work could use selected formatting features, for example to enhance the detection of large homogeneous ranges and headers.

### Combining Spreadsheet Static Analysis with Artificial Intelligence

The technique presented in this paper is one example of how spreadsheet static analysis can be combined with artificial intelligence. Other techniques can also be used to improve the reliability, accuracy, and efficiency of AI-assisted spreadsheet tasks. For instance, (1) formula verification or spreadsheet verification can be applied to AI-generated outputs in order to assess their validity before they are presented to users, or to prompt the AI system to revise them; and (2) tasks or subtasks within a user request—such as formula correctness verification or formula optimization—can be identified and handled by deterministic spreadsheet analyses, while other components are handled by AI. This combination can provide stronger correctness guarantees for the parts amenable to static analysis and help mitigate AI hallucinations.